\documentclass[twocolumn,aps,prb]{revtex4}
\usepackage{graphicx}
\usepackage{bm}
\usepackage{amsmath}
\usepackage{amssymb}

\renewcommand{\vec}[1]{\bm{#1}} 

\newcommand{\tvec}[2]{\begin{pmatrix} #1 \\ #2 \end{pmatrix}}
\begin{document}
\title{
Conductance oscillation due to the geometrical resonance in
$FNS$ double junctions
}
      
\author{Hiroyuki Ohtori$^{1,2}$, Hiroshi Imamura$^{2}$}
\address{
$^1$ 
Institute of Applied Physics, University of Tsukuba, Tsukuba 305-8573, Japan
\\
$^2$NRI-AIST, Central 2, 1-1-1 Umezono, Tsukuba 305-8568, Japan
}

\pacs{}

\begin{abstract}
 We theoretically analyzed the Andreev reflection in ferromagnetic metal /
 nonmagnetic metal / superconductor double junctions with special
 attention to the electron interference effect in the nonmagnetic metal
 layer.
 We showed that the conductance oscillates as a function of the bias
 voltage due to the geometrical resonance. We found that the exchange
 field and therefore the spin polarization of the ferromagnetic metal can be
 determined from the period of the conductance oscillation, which is
 proportional to the square-root of the exchange field.
\end{abstract}
\maketitle

Recently much attention has been focused on
the Andreev reflection (AR) in ferromagnetic metal (FM) / superconductor (SC)
contacts\cite{deJong1995,Soulen1998,Upadhyay1998,Soulen1999,Kikuchi2001,Strijker2001,Ji2001,Mazin2001,Imamura2002,Woods2004}
since the spin polarization of conduction electrons is measured through
the suppression of the conductance below the superconducting gap.
This method is called point contact Andreev reflection (PCAR)
spectroscopy.

 On the other hand, the quasiparticle (QP) interference in
nonmagnetic metal(NM) / SC junctions has
been extensively studied in the past\cite{deGennes1963,Tomasch1965,Tomasch1966,McMillan1966,Rowell1966,Nesher1999,Chang2004,Shkedy2004}.
As shown in Refs. \cite{Tomasch1965,Tomasch1966,McMillan1966}, the
interference of QPs in the SC layer brings about the 
oscillation of the density of states against the QP energy,
which is called a Tomasch oscillation.
The interference in the NM layer also brings about the
oscillation of the density of states in the NM
layer\cite{deGennes1963,Rowell1966}, which is known as the
deGennes-Saint-James bound state or the McMillan-Rowell oscillation.
Nesher and Koren measured the dynamic resistance of
YBa$_{2}$Cu$_{3}$O$_{6.6}$ / YBa$_{2}$Cu$_{2.55}$Fe$_{0.45}$O$_{y}$ /
YBa$_{2}$Cu$_{3}$O$_{6.6}$ junctions and determined the renormalized
Fermi velocity of QPs in the
YBa$_{2}$Cu$_{2.55}$Fe$_{0.45}$O$_{y}$ layer from the period of the
McMillan-Rowell oscillation\cite{Nesher1999}.

In this paper, we theoretically analyze the Andreev reflection in
a FM/NM/SC double junction system with special attention to the electron
interference effect in the NM layer. Following the work of Blonder,
Tinkham, and Klapwijk (BTK) \cite{Blonder1982}, we solve the Bogoliubov-de
Gennes (BdG) \cite{BdG} equations and calculate the conductance.
We show that 
the conductance due to the Andreev reflection oscillates as a function
of the bias voltage because of the geometrical resonance predicted by
deGennes and Saint-James. 
We obtain the analytical expression of the probability of the Andreev
reflection under the Andreev approximation and 
find that the period of the conductance oscillation is proportional to the
square-root of the exchange field.
Therefore, we can determine the exchange field and
therefore the spin polarization of the FM layer from the period of the
conductance oscillation.

\begin{figure}[b]
 \centerline{
\includegraphics[width=0.9\columnwidth]{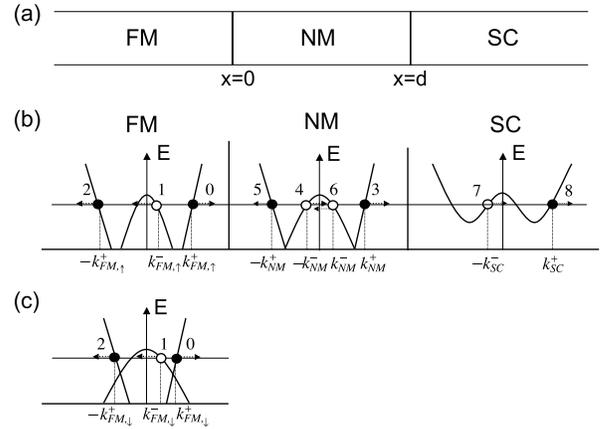}
}
 \caption{
 (a)
 Schematic diagram of a FM/NM/SC double junction. 
 An NM with a thickness of $d$ is sandwiched by FM and SC layers.
 (b)
 Schematic diagrams of energy vs. momentum of the FM/NM/SC double
 junction for a spin-up incident electron are shown.
 The open circles denote holes, the filled circles electrons, and the
 arrows point in the direction of the group velocity.  
 The incident electron with up-spin is denoted by 0,
 along with the resulting scattering processes: Andreev
 reflection (1), normal reflection (2) at the FM/NM interface, 
 transmission to the NM (3, 4) and reflection at the NM/SC interface (5, 6), 
 transmission as a electron-like
 quasi-particle to the SC (7) and that as a hole-like quasi-particle
 (8).
 (c) 
 Schematic diagrams of energy vs. momentum in the FM layer for a
 spin-down incident electron are shown.
 }
 \label{fig:schem}
\end{figure}
The system we consider is comprised of the FM/NM/SC double junctions shown in
Fig. \ref{fig:schem}(a). The current
flows along the $x$-axis, and the interfaces between FM/NM and
NM/SC are located at $x=0$ and $x=d$, respectively.
The system is described by the following BdG
 equation \cite{BdG}:
\begin{equation}
 \begin{pmatrix}
H_0  \!-\!  h(x)\sigma & \Delta (x)  \\
\Delta ^* (x) & \!-\! {H_0 \!-\!  h(x)\sigma}
 \end{pmatrix}
 \begin{pmatrix}
f_{\sigma}(\vec{r})  \\
g_{\sigma}(\vec{r}) 
 \end{pmatrix}
\! = \!
E
 \begin{pmatrix}
f_{\sigma}(\vec{r})\\
g_{\sigma}(\vec{r})
 \end{pmatrix},
 \label{eq:BdG}
\end{equation}
where $H_0 \equiv -({\hbar}^2/2m){\nabla}^2 + V(x) -\mu_F$ is the single
particle Hamiltonian, $E$ is the QP energy measured from the Fermi
energy $\mu_F$, $V(x)$ is the interfacial barrier\cite{tagirov2006}, and
$\sigma=+(-)$ represents the up-(down-)spin band.  
The exchange field function $h(x)$ is given by 
$ h(x)= h_{0} [1-\Theta (x)]$
where $h_{0}$ represents the exchange field in the FM
layer and $\Theta (x)$ is the Heaviside step function.
We employed the two-band Stoner model for the FM layer for simplicity.
The superconducting gap function is expressed as
$\Delta (x) = \Delta_{0}\Theta(x-d)$,
where $\Delta_0$ represents the superconducting gap in the SC layer.
We assume that the system has
translational symmetry in the transverse ($y$ and $z$) direction, and
therefore the wave vector parallel to the interface
$\vec{k}_{\parallel}\equiv(k_{y},k_{z})$ is a conserved quantity.

The general solutions of the BdG equation \eqref{eq:BdG} in the
FM (NM) layer are given by
\begin{align}
 & \Psi _{ \pm k_{\rm FM (NM),\sigma}^+ }(\vec{r})
 = \tvec{1}{0} 
 e^{ \pm ik_{\rm FM (NM),\sigma}^+ x} 
 \, {\rm S}_{\vec{k}_{\parallel}}(\vec{r}_{\parallel}),
 \\
 & \Psi _{ \pm k_{\rm FM (NM),\sigma}^- }(\vec{r})
 = \tvec{0}{1}
 e^{ \pm ik_{\rm FM (NM),\sigma}^- x} 
 \, {\rm S}_{\vec{k}_{\parallel}}(\vec{r}_{\parallel}),
 \label{eq:wf-FM}
\end{align}
where ${\rm S}_{\vec{k}_{\parallel}}(\vec{r}_{\parallel})$ represents the
eigen function in the transverse direction in the $\vec{k}_{\parallel}$
channel and 
$k_{{\rm FM (NM)}, \sigma}^{+(-)}$ is the $x$ component of the wave number
of an electron (hole) with $\sigma$-spin defined as
$
 k_{\rm FM,\sigma}^{\pm} 
= \frac{\sqrt{2m}}{\hbar} \sqrt{\mu _F \pm E + \sigma h_{0} - E_{\parallel}}
$
and 
$
 k_{\rm NM}^{\pm} 
= \frac{\sqrt{2m}}{\hbar} \sqrt{\mu _F \pm E - E_{\parallel}},
$
where $E_{\parallel} = \frac{\hbar^{2}}{2m}k_{\parallel}^{2}$.
In the SC layer, we have
\begin{align}
 & \Psi _{ \pm k_{\rm SC}^+ } (\vec{r})
 = \tvec{u_{0}}{v_{0}}
 e^{ \pm ik_{\rm SC}^+ x} 
 \, {\rm S}_{\vec{k}_{\parallel}}(\vec{r}_{\parallel}),
 \\
 & \Psi _{ \pm k_{\rm SC}^- } (\vec{r})
 = \tvec{v_{0}}{u_{0}}
 e^{ \pm ik_{\rm SC}^- x}
 \, {\rm S}_{\vec{k}_{\parallel}}(\vec{r}_{\parallel}),
 \label{eq:wf-SC} 
\end{align}
where $u_{0}$ and $v_{0}$ are the coherence factors expressed as
$ u_{0}^2  = 1 - v_{0}^2  = \frac{1}{2}\left[ {1 + \frac{{\sqrt {E^2  - \Delta ^2 } }}{E}} \right] ,
$
and $k_{\rm SC}^{+(-)}$ is the $x$ component of the wave number
of an electron-(hole-)like QP defined as
$
 k_{\rm SC}^{\pm} = \frac{\sqrt{2m}}{\hbar}  \sqrt{\mu _F \pm \sqrt{E^2 - \Delta^2
}-E_{\parallel}}.
$

The wave function of the FM/NM/SC double junction is given by
the linear combination of the above general solutions.
Let us consider the scattering of an electron 
in the $\vec{k}_{\parallel}$ channel with $\sigma$-spin injected
into the NM from the FM, the eight processes shown in
Fig.\ref{fig:schem} (b) are active. Therefore, the wave function in the
FM layer $(x<0)$ takes the form
\begin{equation}
 \begin{aligned}
&\Psi ^{\rm FM}_{\sigma,\vec{k}_{\parallel}}(\vec{r})
= \Bigg[ 
\tvec{1}{0}
e^{ik_{\rm FM,\sigma}^{+}x}
+ a_{\sigma,\vec{k}_{\parallel}} \tvec{0}{1}
e^{ik_{\rm FM,\sigma}^{-} x}
\\
&
+ b_{\sigma,\vec{k}_{\parallel}}
\tvec{1}{0}
e^{-ik_{\rm FM,\sigma}^{+} x}
\Bigg] {\rm S}_{\vec{k}_{\parallel}}(\vec{r}_{\parallel}).
\\\\\label{eq:wf2-FM1}
 \end{aligned}
\end{equation}

In the NM layer $(0 \le x < d)$, we have
\begin{equation}
 \begin{aligned}
&\Psi ^{\rm NM}_{\sigma,\vec{k}_{\parallel}}(\vec{r})
  = \Bigg[
\alpha_{\sigma,\vec{k}_{\parallel}} 
\tvec{1}{0}
e^{ik_{\rm NM}^+ x}
+ 
\beta_{\sigma,\vec{k}_{\parallel}} 
\tvec{0}{1}
e^{-ik_{\rm NM}^- x}
\\
&
+ \xi_{\sigma,\vec{k}_{\parallel}} 
\tvec{1}{0}
e^{-ik_{\rm NM}^+ x} 
+ \chi_{\sigma,\vec{k}_{\parallel}} 
\tvec{0}{1}
e^{ik_{\rm NM}^- x}
\Bigg]
{\rm S}_{\vec{k}_{\parallel}}(\vec{r}_{\parallel}),
\label{eq:wf2-SC}
 \end{aligned}
\end{equation}
and in the SC layer $(x \ge d)$
\begin{equation}
\begin{aligned}
&\Psi ^{\rm  SC}_{\sigma,\vec{k}_{\parallel}}(\vec{r})
=
\Bigg[
c_{\sigma,\vec{k}_{\parallel}} 
\!
\tvec{u_{0}}{v_{0}}
e^{ik_{\rm SC}^{+} x}
\\
 &
+ d_{\sigma,\vec{k}_{\parallel}}
\tvec{v_{0}}{u_{0}}
e^{-ik_{\rm SC}^{-} x}
\Bigg] 
{\rm S}_{\vec{k}_{\parallel}}(\vec{r}_{\parallel}).
\label{eq:wf2-FM2}
 \end{aligned}
\end{equation}
The coefficients $a_{\sigma,\vec{k}_{\parallel}}$,$b_{\sigma,\vec{k}_{\parallel}}$,$c_{\sigma,\vec{k}_{\parallel}}$,
$d_{\sigma,\vec{k}_{\parallel}}$,$\alpha_{\sigma,\vec{k}_{\parallel}}$,$\beta_{\sigma,\vec{k}_{\parallel}}$, 
$\xi_{\sigma,\vec{k}_{\parallel}}$, and $\chi_{\sigma,\vec{k}_{\parallel}}$ are
determined by matching the 
wave function at the boundary of the contact $x=0$ and $d$.
Following the BTK theory\cite{Blonder1982} the probabilities of the AR
and the normal reflection are given by
$A_{\sigma,\vec{k}_{\parallel}}(E) = (k_{\rm FM,\sigma}^-/k_{\rm
FM,\sigma}^+) a_{\sigma,k_{\parallel}}^* a_{\sigma,k_{\parallel}}$
and
 $B_{\sigma,\vec{k}_{\parallel}} (E) = b_{\sigma,k_{\parallel}}^*
 b_{\sigma,k_{\parallel}}$, respectively.
Since we assume that the temperature is zero, the conductance at bias
voltage $V$ is given by 
$
 G  = \frac{e}{h}\sum_{\sigma,\vec{k}_{\parallel}}
\left[
 1 + A_{\sigma,\vec{k}_{\parallel}}(eV) - B_{\sigma,\vec{k}_{\parallel}}(eV)
\right],
$
where we assume that the voltage drop occurs at the NM/SC interface for
simplicity.
Below the superconducting gap, i.e., $eV < \Delta_{0}$, 
the probabilities $A_{\sigma,\vec{k}_{\parallel}}(E)$ and
$B_{\sigma,\vec{k}_{\parallel}}(E)$ satisfy the relation that
$
 1 - B_{\sigma,\vec{k}_{\parallel}}(E) = A_{\sigma,\vec{k}_{\parallel}}(E)
$
and then we have
\begin{equation}
 G  = 2\frac{e}{h}\sum_{\sigma,\vec{k}_{\parallel}} A_{\sigma,\vec{k}_{\parallel}}(eV).
\label{eq:gA}
\end{equation}

 Let us first consider the most idealistic case where the interfacial
 scattering potential, $V(x)$, is assumed to be zero.  This
 simplification enables us
 to obtain the analytical expression of
 $A_{\sigma,\vec{k}_{\parallel}}(E)$ under the Andreev approximation.
\begin{widetext}
\vspace{-2ex}
 \begin{equation}
A_{\sigma,\vec{k}_{\parallel}}(E) 
 \simeq
 \frac{4(1\! - \!\zeta)\sqrt{1\! - \!\eta^{2}}}{
 2\left[(1\! - \!\zeta)^2 \! +\! (1\! - \!\zeta)\sqrt{(1\! - \!\zeta)^{2}\! - \!\eta^{2}})\right]
 \! - \! \eta^{2}
 \left[
  \epsilon^{2} 
  \! +\! (1\! - \!2\epsilon^{2})\cos^{2}[(k_{\rm NM}^{+}\! - \!k_{\rm NM}^{\! - \!})d]
  \! - \!\epsilon\sqrt{1\! - \!\epsilon^{2}}\sin[2(k_{\rm NM}^{+}\! -
  \!k_{\rm NM}^{-})d]
 \right]
 },
 \label{eq:prob3d}
 \end{equation}
\end{widetext}
where we introduced the normalized parameters $\eta=h_{0}/\mu_{\rm F}$,
$\zeta=E_{\parallel}/\mu_{\rm F}$, and $\epsilon=E/\Delta_{0}$.
We note that Eq. \eqref{eq:prob3d} contains trigonometric functions
in the denominator. For the FM/SC junction, i.e., $d$=0, the
trigonometric functions become constant and Eq. \eqref{eq:prob3d}
reproduces the deJong and Beenakker results of the zero-bias
conductance\cite{deJong1995}.
For the FM/NM/SC junctions with $d\neq 0$, the trigonometric functions 
in Eq. \eqref{eq:prob3d} give rise to the oscillation of the conductance
against the bias voltage. The origin of the conductance oscillation
is the interference of electrons in the NM layer \cite{deGennes1963,Rowell1966}.
In the FM/NM/SC double junctions, the injected electron propagates across
the NM layer to the interface as an electron (3 in
Fig.\ref{fig:schem}) and is scattered into a hole (6 in
Fig.\ref{fig:schem}) by the superconducting gap.
The superconducting gap can pair an excited electron with an
electron inside the Fermi sea, leaving a hole excitation. The
hole propagates back across the NM layer; however, it cannot interfere
with the original electron. In
order for interference to occur the hole must be reflected at the FM/NM
interface, propagate to the NM/SC interface, be scattered into the electron
state (5 in Fig.\ref{fig:schem}) by the superconducting gap, and
propagate as an electron in the NM layer. It can interfere with the
original electron (3 in Fig.\ref{fig:schem}).
This interference produces an oscillation of the conductance against
the bias voltage, and the period of the oscillation is determined
by the thickness of the NM layer.

\begin{figure}[t]
 \centerline{
  \includegraphics[width=\columnwidth]{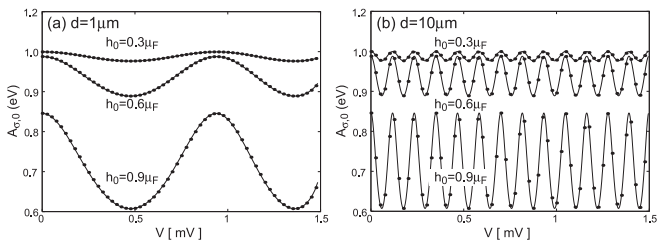}
 }
 \caption{
 (a) Probability of the AR $A_{\sigma,0}(eV)$ for the
 FM/NM/SC double junction with $d$=1$\mu$m is plotted against the
 bias voltage $V$. The exact numerical results and the value of
 Eq. \protect\eqref{eq:prob3d} are plotted by lines and circles, respectively.
 The value of the exchange field is taken to be $h$ = 0.3, 0.6, 0.9 $\mu_{\rm
 F}$ from top to bottom.
 (b) Same plot for $d$=10$\mu$m.
 }
 \label{fig:1d}
\end{figure}

In order to analyze the interference effect on the conductance
oscillation, we consider the AR probability
$A_{\sigma,\vec{k}_{\parallel}}(E)$ of the 
one-dimensional system; i.e., only the transverse channel with
$\vec{k}_{\parallel}=0$ is considered.
In Fig. \ref{fig:1d} (a), \ref{fig:1d} (b) we plot the 
probability $A_{\sigma,\vec{k}_{\parallel}}(eV)$
of the FM/NM/SC double junctions with $d$=1$\mu$m and 10$\mu$m,
respectively, as a function of the bias voltage. 
Since Eq.\eqref{eq:prob3d} is an even function of normalized value of
the exchange field $\eta$, $A_{\sigma,\vec{k}_{\parallel}}(eV)$ is
independent of the spin direction $\sigma$; i.e.,
$A_{\uparrow,\vec{k}_{\parallel}}(eV) =A_{\downarrow,\vec{k}_{\parallel}}(eV)$.
The Fermi energy and the superconducting gap are assumed to be
$\mu_{F}=3.8$ eV $(k_{F}= 1.0 {\rm \AA}^{-1})$ and $\Delta_{0}=1.5$meV,
respectively.
The exact numerical results and the approximate
values of Eq. \eqref{eq:prob3d} are
plotted by lines and circles, respectively. The value of the exchange
field is taken to be $h_{0}$ = 0.3, 0.6, 0.9 $\mu_{\rm F}$ from top to
bottom.

As shown in Figs. \ref{fig:1d} (a) and \ref{fig:1d} (b),
Eq. \eqref{eq:prob3d} and
therefore the Andreev approximation are valid for all values of the
exchange field. According to Eq. \eqref{eq:prob3d}, the period of the
oscillation is determined by the condition that $k^{+} -  k^{-} = n
\pi$, where $n$ is an integer. Since $k^{\pm} \simeq k_{\rm F}(1 \pm
E/2\mu_{\rm F})$, the period is obtained as
\begin{equation}
\Delta V_{\rm 1D} \simeq  \frac{\hbar^{2}\,\pi\, k_{\rm F}}{2m\, e\, d},
\label{eq:v1d}
\end{equation}
which is inversely proportional to the thickness of the NM layer, $d$.
For the one-dimensional system, the period $\Delta V_{\rm 1D}$ is
independent of the exchange field of the FM layer as shown in
Figs. \ref{fig:1d} (a) and \ref{fig:1d} (b), and in Eq.\eqref{eq:v1d}.
However, as we shall show later, the period of the conductance
oscillation due to the geometrical resonance depends
on the exchange field of the FM layer because the number of
$\vec{k}_{\parallel}$ channels available for the AR is
restricted by the exchange field.

The period of the oscillation of $A_{\sigma,\vec{k}_{\parallel}}(eV)$
with finite $\vec{k}_{\parallel}$ is given by
$\Delta V_{\vec{k}_{\parallel}} 
 \simeq
 \frac{\hbar \pi }{\sqrt{2m}ed}
 \sqrt{\mu_{\rm F} - E_{\parallel}}$.
From Eq.\eqref{eq:gA} the conductance is obtained by summing up 
$A_{\sigma,\vec{k}_{\parallel}}(eV)$ for all available
$\vec{k}_{\parallel}$. Since the spin of the Andreev reflected hole is
opposite to that of the incident electron, the maximum value of
$k_{\parallel}$ and therefore $E_{\parallel}$ is limited by the exchange
field $h_{0}$ as $\max{E_{\parallel}} = \mu_{\rm F} - h_{0}$.
We assume that oscillations of $A_{\sigma,\vec{k}_{\parallel}}(eV)$ 
with different periods cancel out each other and 
the period of the sum $\sum_{\sigma,\vec{k}_{\parallel}}
A_{\sigma,\vec{k}_{\parallel}}(eV)$ is determined by the shortest period.
Thus, the period of the conductance oscillation of the three-dimensional
system is obtained as
\begin{equation}
\Delta V_{\rm 3D} \simeq \min{\Delta V_{\vec{k}_{\parallel}}} = 
 \frac{\hbar\, \pi}{\sqrt{2m}\, e \,d}
 \sqrt{h_{0}}.
\label{eq:v3d}
\end{equation}

\begin{figure}[b]
\centerline{
  \includegraphics[width=\columnwidth]{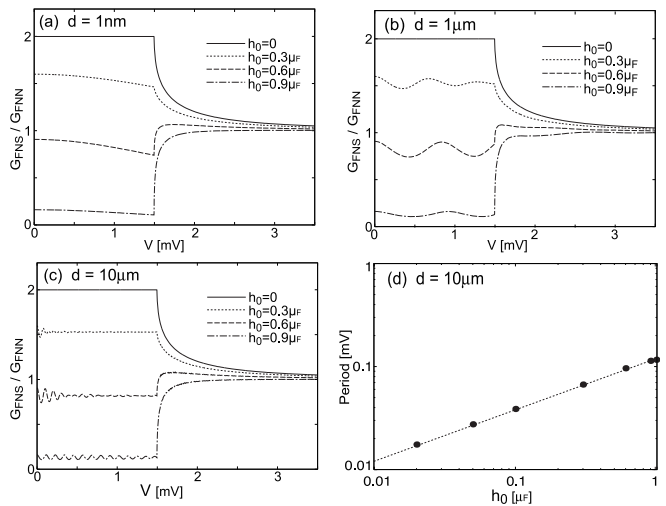}
 }
 \caption{
 (a) The conductance of the FM/NM/SC double junction, $G_{\rm FNS}$ with $d$=1nm is plotted
 against the bias voltage. The conductance is normalized by that of the
 FM/NM/NM junction, $G_{\rm FNS}$.
 (b) Same plot for $d$=1$\mu$m.
 (c) Same plot for $d$=10$\mu$m.
 (d) The period of the conductance oscillation of the FM/NM/SC double
 junction with $d$=10$\mu$m is plotted against the exchange field $h_{0}$.
 }
 \label{fig:3d}
\end{figure}

In Figs. \ref{fig:3d} (a), \ref{fig:3d} (b), and \ref{fig:3d} (c) we
plot the conductance of the FM/NM/SC junction,$G_{\rm FNS}$,
normalized by that of the FM/NM/NM junction, $G_{\rm FNN}$, against the
bias voltage. As shown in Fig. \ref{fig:3d} (a), the oscillation due to the
geometrical resonance does not appear in the conductance-voltage curve
if the thickness of the NM layer, $d$, is less than or of the order of
nm. The conductance-voltage curve is indistinguishable from that of the
FM/SC junction. Hence, we can use the conventional PCAR analysis
for a FM film, the surface of which is coated by a thin (less than a few
nm) NM layer.

Figures \ref{fig:3d} (b) and \ref{fig:3d} (c) show the
conductance-voltage curves for the FM/NM/SC double junctions with $d=1\mu$m
and $d=10\mu$m, respectively. One can see that the period does depend
on the exchange field, $h_{0}$, in the FM layer as well as the
thickness of the NM layer. The period is a decreasing function of the
exchange field. We can easily confirm that the period is proportional
to the square-root dependence of the exchange field by looking at
Fig. \ref{fig:3d} (d). The exact numerical results (filled circles)
agree well with Eq. \eqref{eq:v3d} (dotted line).
The results suggest that we can determine the exchange field and
therefore the spin polarization of the FM layer from the period of the
conductance oscillation.

\begin{figure}[t]
\centerline{
  \includegraphics[width=\columnwidth]{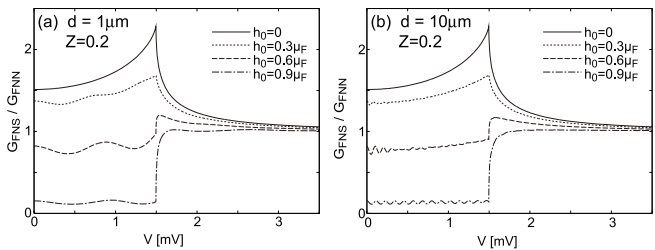}
 }
 \caption{
 (a) The normalized conductance, $G_{\rm FNS}/G_{\rm FNN}$ for
 $d$=1$\mu$m, Z=0.2 is plotted
 against the bias voltage. 
 (b) Same plot for $d$=10$\mu$m, Z=0.2.
 }
 \label{fig:z}
\end{figure}

Next we consider the effect of the interfacial scattering potential at
the NM/SC interface.  We assume that the interfacial potential is
represented by the delta-function as $V(x) = (\hbar^2
k_{F}Z/m)\delta(x-d)$, where $Z$ is the dimension less parameter
which characterize the strength of the interfacial scattering potential.
For simplicity, we neglect resistances of the FM/NM interface and the FM
layer which do not change the oscillation period of the conductance but
reduce the amplitude of it.
In Figs.\ref{fig:z} (a) and \ref{fig:z} (b), we show the normalized
conductance-voltage curves for junctions with $d$=1$\mu$m and
$d$=10$\mu$m. The parameter $Z$ is assumed to be 0.2\cite{Strijker2001}.
Because the conductance oscillation is due to the geometrical resonance
in the NM layer, the
period of the oscillation is not affected by the interfacial scattering
potential and is given by Eq.\eqref{eq:v3d}.

In the present analysis we employed the simplest BdG approach and
consider the clean FM/NM/SC junctions with perfect interfaces.
In the real experiments the conductance
oscillation we predicted might be smeared out due to the interface
roughness and imperfections.
However, recent advances in fabrication technology enables us to
fabricate epitaxial FM/NM/SC trilayers of high
quality\cite{yamazaki2006}.  We expect that the conductance oscillation
we predicted can be observed in such epitaxial trilayers.
For further understanding of the transport properties of FM/NM/SC
trilayers, we have to take into account the effects of finite mean free
path, band structures, and selfconsistent determination of the
electron's distribution function and electric-pottential, which is
beyond the scope of this Brief Report.


In summary, we studied the 
conductance oscillation due to the geometrical resonance in
a FM/NM/SC double junction theoretically. We showed that the conductance due to the
Andreev reflection oscillates as a function of the bias voltage due to
the geometrical resonance. We found that the exchange field and
therefore the spin polarization of the FM layer can be determined from
the period of the conductance oscillation because the period of the
conductance oscillation is proportional to the square-root of the
exchange field.

The authors would like to acknowledge the valuable
discussions they had with K. Matsushita and N. Yokoshi.

\end{document}